# X-optogenetics and U-optogenetics: Feasibility and Possibilities

**Rachel Berry, Matthew Getzin, Lars Gjesteby, and Ge Wang**
**Rensselaer Polytechnic Institute, Troy, NY 12180**

**Abstract**

Optogenetics as developed by Dr. Karl Deisseroth and others has been a transformative technology in the area of neuroscience. By stimulating genetically-modified neurons with visible light, modulation of the ionic conduction across the cell membrane can be achieved with very high specificity and precise temporal control. Despite the major influence of this technique, its scope is limited by its invasive nature and its lack of stimulation depth. Visible light is unable to penetrate deeply into biological soft tissue and has even greater difficulty passing through hard tissues such as bone. Therefore, to deliver the light to the nerves, a window in the subject's skull must be surgically made and a light probe, either LED or laser, is inserted near the area of interest. At this point, the surface of the cortex is available for stimulation as visible light is interacting with the soft tissue of the brain. Recent grants have been awarded to groups that investigate ways to eliminate the need for the invasive surgery by introducing fluorescence emitting nanoparticles into the brain. Through stimulation of these particles with infrared light, subsequent visible fluorescence occurs. Hence, the stimulating light source can be moved into the brain and functionally targeted to the genetically-modified cells. However, despite the incrementally improved depth available with IR stimulation, much of the brain remains out of reach. To address these limitations, we propose two new methods for optogenetic stimulation. The first is x-optogenetics, which uses visible light-emitting nanophosphors stimulated by focused x-rays. This idea is not new but the application to optogenetics is novel. X-rays can penetrate much more deeply than infrared light and could allow for nerve cell stimulation in any part of the brain. In this paper, we discuss the feasibility and possibilities of such a method by describing the advances in nanomaterials, x-ray focusing, and x-ray sources. Also, we discuss concerns when dealing with x-rays such as radiation dosage. Through the use of quantities and assumptions backed by recent literature, manufacturer specifications, and personal correspondence, a full feasibility analysis of x-optogenetics is completed. The second proposed method we explore is u-optogenetics, which is the application of sonoluminescence to optogenetics. Such a technique uses ultrasound waves instead of x-rays to induce light emission, so there would be no introduction of radiation. However, the penetration depth of ultrasound is less than that of x-ray. The key issues affecting feasibility are laid out for further investigation into both x-optogenetics and u-optogenetics.

## 1. Introduction

After a transformative technology is invented, there is often a period of improvements and optimizations that expands the reach of the technology. Optogenetics, an incredibly innovative technology that allows for deep insight in the field of neuroscience and neuropathology, falls into this category of transformative technologies. Since its introduction into mainstream science less than twenty years ago, a number of teams have adopted the technique to study the roles of various neurons in disease states such as Parkinson's, epilepsy, and depression[1]. However, these applications are limited in their scopes because of the invasive nature and depth limitation of optogenetics. There is a critical and immediate need to improve optogenetics for deeper and non-invasive applications.

In this paper, we will briefly review the optogenetics technique focusing on the areas that need improvement, and introduce two possible enhancements that seek to eliminate invasiveness and overcome depth limitations. One of our proposed techniques takes advantage of the recent advances in both nanomaterials and x-ray optics. Through this unique combination, x-optogenetics can deeply target nerves without any surgical intervention. The other method aims to take advantage of methods previously demonstrated *in vivo* that enable ultrasonic-induced luminescence without radiation[2]. Both methods offer transformative improvements to what is already a powerful technology.

## 2. Methodology and Results

### 2.1. Optogenetics

Optogenetics refers to the technology that uses visible light to trigger proteins that modulate membrane potentials in neuronal cells through excitatory or inhibitory membrane currents[3]. This ability of controlling neuronal cells has proven instrumental in preclinical studies and holds enormous potential for the treatment of diseases such as Parkinson's, depression, and vision-impairments[1]. However, the current techniques used for optogenetic control remain too invasive for clinical applications. These techniques are briefly described below to describe the motivation behind our innovative ideas for advancing their efficiency and scope.

Created by Karl Deisseroth, the original form of optogenetics uses channelrhodopsin2 (ChR2) to induce excitatory potentials in transfected neurons of small animals. ChR2 is a protein found in green algae that becomes permeable to cations in the presence of blue light. Deisseroth rationalized that ChR2 could be used in neurons since ion channels are a main contributor in electrical signal transduction in the brain. Since then, the technology has allowed researchers to target specific areas of the brain and study how modulated neuron firing affects downstream behaviors and cellular processes[4,5].

Optogenetics is performed in multiple steps: first, specified neuronal cells are transfected with DNA encoding for light-sensitive ion channels such as ChR2, halorhodopsin (NphR), and archeorhodopsin (Arch). Upon expression of these proteins, scientists surgically implant a light fiber into the organism's brain so light at the stimulating wavelengths can directly irradiate neurons and modulate their membrane current. Membrane current modulation comes in two forms that depend on the ions to which the channel becomes permeable in its open state. The cation specific channels lead to membrane depolarization (excitatory) and the anion specific channels cause the membrane to hyperpolarize (inhibitory)[3].

In this way, the membrane current is directly controlled by a light source which is currently in the form of either a laser or a LED. Both sources have limitations. Lasers are very costly and have some light loss. On the other hand, the light from a LED is spread out and is not a straight beam, so it cannot be accurately targeted as laser can[6–9]. It should be noted that current practice of optogenetics is performed on a macroscopic scale. For example, ChR2-expressing cells have been activated by a 470-490 nm light in power range 1-20 $mW/mm^2$ and pulse duration 5-100 ms. This type of stimulation results in ChR2-channel driven membrane current that peaks around -9 pA/pF[6–8,10]. In these experiments, the studied tissue is flooded with light and any cell within a few millimeters of the source that is expressing light-sensitive ions channels will have modulated membrane currents that may lead to distinct network and/or behavioral changes. The proposed x-optogenetic and u-optogenetic techniques shift the scale from the macroscopic to a microscopic or even nanoscopic level of control. This will become more apparent in the following sections and the feasibility analysis.

Because of the invasive nature or limited penetration of LEDs and laser sources, researchers are trying to find light sources that do not require surgically implanting a probe into an organism and that deliver light more deeply. For example, Dr. Gang Han uses near infrared radiation to excite upconversion nanoparticles (UCNPs). After excitation these nanoparticles will emit photons of light whose wavelengths can be customized based on the particle chemistry. These emissions are then used to modulate the membrane current just as the light sources described above. This method offers a unique and less invasive approach to optogenetics as deeper levels of the brain can be mapped due to the deeper penetrating abilities of infrared light[11]. Table 1 lists a number of these light-emitting nanoparticles (nanophosphors) which have been reported in the literature and may have utility in this regard. Despite this incremental improvement to the optogenetic approach, infrared (IR) light has its limitations as well. First, IR penetration through the skull has been shown to be between 4 and 10% of the initial intensity[12]. Furthermore, IR light at 868 nm only penetrates brain tissue around 2.5 mm[13], which can only gain access to a fraction of human cortical neurons as the human cortex thickness is typically on the range of 2 to 5.5 mm[14]. We believe that these challenges can be addressed through the use of x-rays, rather than infrared light, to stimulate nanophosphors. X-rays are capable of very deep penetration and can also be precisely focused as is discussed below. Due to the radiation dose introduced by x-rays, we also explore the possibility of ultrasonic stimulation of air bubbles that emit light via sonoluminescence. This technique would not introduce radiation to the subject, but the penetration depth would not be as high as that of x-ray stimulation methods.

## 2.2. Key Elements for X-optogenetics

### 2.2.1. X-ray Excitable Nanophosphors

To perform x-optogenetics, the x-rays must be converted to visible light; therefore, x-ray excitable nanophosphors must be used. The nanophosphors need to be biocompatible and emit light at wavelengths that properly activate the light sensitive ion channels. This idea was recently used as the basis of a patent application, though it simply describes using x-ray excitable nanophosphors for general "control of light-sensitive bioactive molecules" without any quantitative analysis of feasibility or radiation dose implications[15]. A literature survey of nanophosphors verified that a large number of nanophosphors can be readily produced with tunable emission, absorbance, and solubility properties. Table 1 shows a large number of nanophosphors with emission maxima in the visible domain. With regards to their excitation spectra, however, two distinct types of nanoparticles can be seen: up-conversion nanoparticles (UCNPs) and UV/x-ray excitable nanoparticles. As stated above, UCNPs emit visible photons during exposure to long wavelength infrared radiation, while the UV/x-ray excitable particles emit visible photons during exposure to short wavelength UV/x-ray radiation. It should be noted that particles in the same conversion class are often doped with similar ions. For example, UCNPs often contain $Yb^{3+}$, $Ln^{3+}$, or Er. On the other hand, particles sensitive to the shorter wavelength radiation often contain $Cr^{3+}$, $Eu^{3+}$, or $Tb^{3+}$.

Of the reported nanophosphors, most have the excitation wavelength in the range from 147nm to 980nm. There are five nanophosphors in the survey with an excitation wavelength of 980nm all of which are upconverting nanoparticles (UCNPs). X-rays have a wavelength range from 0.01 to 10 nm[16] and therefore cannot efficiently excite these nanophosphors. However, the particles with the base chemistry of $Gd_2O_2S$ and $LiGa_5O_8$ have been shown to absorb light in both the UV and x-ray ranges. Additionally, these particles have been doped with $Cr^{3+}$, $Eu^{3+}$, or $Tb^{3+}$. Other particles in the survey also utilize these dopants and may also prove to be useful for x-ray excitation. Further research should generate optimal nanophosphors for x-optogenetics.

The results from the literature suggests that there are a number of techniques available to improve the nanophosphors that can be used for x-optogenetics, especially in the areas of solubility, conversion efficiency, emission, size, and targeting. For example, Table 1 includes nanophosphors that emit light across the visible light spectrum and into the NIR range with the shortest wavelength emitted at 450 nm and the longest at 800nm. The emission wavelength is a result of the chemical formula of the nanophosphor and the compound with which it is doped. For example, $NaYF_4$ doped with $Eu^{3+}$ has an emission wavelength of 592 nm while $NaYF_4$ doped with $Tb^{3+}$ has an emission wavelength of 545 nm. The ability to alter a nanophosphor's emission wavelength by changing the chemical formula or the compound with which it is doped should be beneficial in optimizing nanophosphors for x-optogenetics. Hybrid doping schemes may also allow for more tailored emission spectra.

Table 1. Survey of nanophosphors in the literature. Italicized entries represent nanophosphors which may be useful for x-optogenetics as $Eu^{3+}$ and $Tb^{3+}$ are common dopants for x-ray excitable nanophosphors. Not all of these particles reported x-ray induced fluorescence. (PEG – polyethylene glycol, PAA – poly(acrylic acid), PGA – polyglycolic acid, PEI – polyethylenimine, DSPE-PEG-COOH – 1,2-distearoyl-sn-glycero-3-phosphoethanolamine-N-[carboxy(polyethylene glycol)]).

| Formula | Source | Emission Maximum (nm) | Excitation Wavelength (nm) | Conversion Efficiency (%) | Size (nm) | Dispersible | Toxicity |
|---|---|---|---|---|---|---|---|
| $Gd_2O_2S{:}Eu^{3+}$ ($Tb^{3+}$) | 17–20 | *620 (545)* | *<310* | *15* | *50-300* | *Yes, PGA-PEG* | *Low* |
| $Y_2O_3{:}Eu^{3+}$ | 21–23 | *610* | *<310* | *80* | *10-50* | *Yes* | -- |
| $LiGa_5O_8{:}Cr^{3+}$ | 24,25 | *716* | *<310* | -- | *50-150* | *Yes, PEI* | *Low* |
| $Gd_2O_2S$:Yb(8),Er(1) | 26 | 500-700 | 980 | 25 | 4 µm | Yes | Low |
| $NaMF_4$ :$Yb^{3+}/Ln^{3+}$ | 27 | 510-560 | 980 | -- | 60 | Yes, DSPE-PEG-COOH | Low |
| $La(OH)_3{:}Eu^{3+}$ | 28 | *597, 615* | *280* | -- | *3.5* | *Yes, PEG* | *Low* |
| $NaYF_4$:Yb/Er | 29 | 520, 540, 654 | 980 | -- | 33 ± 1 | Yes, citrate | -- |
| $NaYF_4{:}40\%Eu^{3+}$ | 30 | *592* | *394* | -- | *28* | *Yes, PAA* | *Low* |
| $NaYF_4{:}40\%Tb^{3+}$ | 30 | *545* | *368* | -- | *28* | *Yes, PAA* | *Low* |
| cit-$NaLuF_4$:Yb,Tm | 31 | 800 | 980 | -- | 25 | Yes, citric acid | Low |
| $Ba_2SiO_4$ | 32 | 505 | 350 | 38.6 | 40-50 | -- | -- |
| $Na_2Sr_2Al_2PO_4F_9{:}Eu^{3+}$ | 33 | *593, 619* | *393* | -- | *35.26* | -- | *Non-toxic materials* |
| $BaMgAl_{10}O_{17}{:}Eu^{2+}$ | 34 | *450* | *147* | -- | *62, 85, 115, 160, 450* | -- | -- |
| $Sr_2CeO_4$ | 35 | 467-485 | 240-360 | -- | 45 | -- | -- |
| $LiCaPO_4{:}Eu^{2+}{}_{0.03}$ | 36 | *476* | *375* | *Quantum Efficiency: 53.7, 67.6* | | *Yes, PEG-P* | -- |
| PEG-Er-$Y_2O_3$ | 37 | 660 | 980 | -- | 30-60 | Yes, PEG | Low |
| $GdVO_4{:}Eu^{3+}$ | 38 | *620* | *330* | -- | *6* | *Yes* | *Low* |

The conversion efficiency is the ability for the nanophosphors to convert x-ray energy to visible light energy. This value was not expressed for many of the nanophosphors, though it remains an important consideration for x-optogenetic applications. When choosing a nanophosphor for x-optogenetics, the conversion efficiency should be as high as possible to reduce the amount of time and x-ray dose to which the subject is exposed. It is underlined that we consider x-ray stimulation feasible and safe, given the extensive research on x-ray luminescence imaging in preclinical applications[39].

When considering x-optogenetics for neuronal intervention, the size distribution and coating of the nanophosphors is important for penetration of the phosphors across the blood brain barrier (BBB) to gain access to the cells in the brain. Size distribution of particles targeted outside of the central nervous system do not need to be as small, but should still be optimized for maximum bioavailability. Studies using polysorbate-coated nanoparticles showed maximum passage through the BBB for nanoparticles under 100 nm in diameter[40]. With the sizes of the nanophosphors in the survey between 10 nm and 1 µm and recent advances in nanotechnology, it should be feasible to obtain nanophosphors with an appropriate size distribution for a range of x-optogenetic applications[41].

In addition to size, the ability for the particles to be soluble or colloidal in water is a critical property of the nanophosphors as this should add to their biocompatibility. A number of surface coatings including polyethylene glycol (PEG) and other forms of hydrophilic polymers were used to help solubilize or suspend the surveyed particles in aqueous solutions. These types of coatings could be used for the nanophosphors to facilitate x-optogenetics. These coatings can also have a profound effect on the ability of the particles to cross the blood brain barrier[40].

#### 2.2.2. Nanoparticle Targeting

One important issue with x-optogenetics is the placement of the light sources that will be used to generate membrane current in the target neurons. The proximity of these nanophosphors in relation to the light-sensitive ion channels must be within a few millimeters as power density is reduced by >90% after 1 mm for all wavelengths of visible light[42].

One way of combating this light loss through tissue would be to directly target the light-sensitive ion channels through functionalization of the nanoparticles. Several groups have demonstrated the ability to conjugate small peptide sequences or antibodies that can be used to enhance cellular uptake or adhesion to the cellular membrane[43–45]. Using similar methods, the nanophosphors could be functionalized to specifically bind to the light-sensitive ion channels that are on the target neurons. In this way, the proximity issue between the light-sensitive ion channels and the light sources can be minimized and the light loss due to tissue absorption mitigated.

#### 2.2.3. X-ray Focusing

Targeting the genetically-modified neurons through functionalization of the nanophosphors will provide the first level of control for neuron activation. A second level of control comes from the ability to focus the x-rays through the use of a polycapillary lens, a zone plate, or another similar means such as a grating. In addition to enhanced control over the neuronal activation, focused x-rays will result in less bulk x-ray dose to the patient which is always of high concern when dealing with ionizing radiation.

A polycapillary lens focuses x-rays in the form of an intense microspot using an array of glass micro-capillaries. The size of the focal spot can get as low as 5 µm[46]. However, for single neuron targeting focal spots of a few 100 µm may be more applicable. Conventional polycapillary lenses have a working energy range of 0.5 to 30 keV. These can be described as soft x-rays and more easily absorbed by the brain tissues. Also, polycapillary optics has recently been made for focusing of higher energy x-rays up to 60 keV, although transmission through these lenses is <5% at energies higher than 5 keV[47].

Excitation of x-ray excitable nanophosphors using the same mechanisms has been previously proposed and simulated by our group[48]. As described for x-ray fluorescence computed tomography (XFCT) applications, the x-ray intensity distribution in biological soft tissues can be approximated with inverse distance weighting. In this approximation $I(\boldsymbol{r}) = I_0 W(\boldsymbol{r}, \boldsymbol{r}_0)/\|\boldsymbol{r} - \boldsymbol{r}_0\|^2$, where $\mathbf{r}_0$ is the vertex of the double cones, $I_0$ is the intensity of the x-ray source, and $W(\mathbf{r},\mathbf{r}_0)$ is the aperture function of the double cones at the vertex $\mathbf{r}_0$. For accurate membrane current modulation of the target neurons, the initial intensity of the x-ray source can be adjusted so that the nanophosphors near $\mathbf{r}_0$ will receive enough x-ray energy to emit a sufficient number of light photons to open the ion channels. This real-time adjustment is also dependent on the location of the target neurons as well as the size and fluorescence conversion efficiency of the nanoparticles.

Focusing x-rays can be more precise through the use of a Fresnel zone plate (FZP). FZPs are micro-fabricated from a soft metal such as gold or nickel, and modulate either amplitude or phase-shift of incoming x-rays. These modulations result in a wave diffraction and constructive interference at a focal point[49]. One consideration is that zone plates are typically used for synchrotron radiation produced x-rays. Similar to the polycapillary lens, zone plates are most effective for x-rays with lower energy levels (5-8 keV)[49]. All things considered, the polycapillary lens may be initially the best option for x-optogenetics.

#### 2.2.4. X-ray Carbon Nano-Tube (CNT) Sources

Another key aspect of optogenetics that must be addressed by x-optogenetics is the delivery of light in pulses 5 – 100 msec in duration. With the light emitted from the nanophosphors as they are excited by x-rays, the pulsation must come from the x-ray source itself. Conventional tubes emit x-rays under 10 – 500 mA current and require several minutes for warming-up before emission. Achieving a sufficient pulsing emission rate will not be possible with such a source. Fortunately, recent progress has been made on the development of carbon-nanotube field-emission cathodes that can produce soft x-rays at about 8 keV and are capable of pulsing at high rates for x-optogenetics research and application[50].

#### 2.2.5. X-ray Dose

With the involved ionizing radiation for x-optogenetics, it is important to quantify the delivered radiation dose during a procedure. While everyone is subject to a baseline effective dose of about 3 mSv a year, increased levels of radiation exposure occur as a result of x-ray, CT, and/or PET imaging exposure. The effective dose from such a scan can range anywhere from 0.001 mSv to 25 mSv. These values depend on the region of exposure, type of radiation, and type of scan. For x-ray related scans, a highest effective dose administered is around 10 mSv[51]. We will use this number as the highest effective dose permissible for x-optogenetics protocols in the following feasibility analysis.

### 2.3. X-Optogenetics Feasibility and Safety Analysis

Table 2 summarizes the requirements used for the various light-sensitive ion channels in past studies. As stated previously these techniques take a macroscopic view on the requirements of optogenetics. When considering the requirements needed to perform x-optogenetics, a micro/nanoscopic scale must be used and converted to macroscopic changes. The nanophosphors must be excitable with x-rays, be biocompatible, and have a high conversion efficiency. Depending on the nanophosphors used, the x-ray dose may be able to be adjusted. In any case, the nanophosphors should emit visible light or similar photons that can be used for optogenetics. A CNT, polycapillary lens, Fresnel zone plate or a similar component should be used to deliver x-rays.

Table 2. X-optogenetic overview for multiple light-sensitive ion channels. Included in this table are approximate sizes of the channels which helps validate close proximity of nanophosphors and channels after targeting. Furthermore, nanoparticles that can be used for targeting the ion channels are specified.

| Ion Channel | Channelrhodopsin 2 (ChR2) | Halorhodopsin (NphR) | Archeorhodopsin (Arch) |
|---|---|---|---|
| Channel Mass | 60-70 kDa [52] | 30 kDa [53] | 28 kDa [54] |
| Minimum Channel Radius (assuming spherical) | 2.58-2.72 nm [55] | 2.05 nm [55] | 2.00 nm [55] |
| Intensity | 2-20 $mW/mm^2$ [56] | 7.9 $mW/mm^2$ [57] | 76.1 $mW/mm^2$ [57] |
| Wavelength | 488 nm [56] | 532 nm [57] | 532 nm [57] |
| Pulse Train | 5 msec, 40 Hz [56] | 15 sec illumination [57] | 15 sec illumination [57] |
| Depolarizing/Hyperpolarizing | Depolarizing | Hyperpolarizing | Hyperpolarizing |
| Possible Nanophosphors | *$BaMgAl_{10}O_{17}:Eu^{2+}$*<br>*$LiCaPO_4:Eu^{2+}_{0.03}$* | *$Gd_2O_2S:Tb^{3+}$*<br>*(combination doping?)* | *$Gd_2O_2S:Tb^{3+}$*<br>*(combination doping?)* |
| Hardware Specifications/Involved Components | SOURCE: Carbon Nanotube (peak ~8 keV, pulsing capability)<br>FOCUSING ELEMENT: polycapillary lens OR Fresnel zone plate | | |

Assuming a maximum effective radiation dose of 10 mSv, a theoretical calculation of power emitted from the nanophosphors can be done. For x-ray radiation, a Sievert (Sv) is defined as 1 Joule (J) of energy per kilogram (kg) of tissue. By definition, 1 J is equal to 6.24 E 12 MeV[58]. Furthermore, the density of brain tissue is given in the literature to be 1.04 g/cm$^3$ [59]. Using these relationships, the following conversion can be done.

$$0.010\, Sv \left(\frac{1\frac{J}{kg}}{1\, Sv}\right)\left(\frac{6.24E12\, MeV}{1\, J}\right)\left(\frac{0.00104\, kg}{1\, cm^3}\right)\left(\frac{1\, cm}{10\, mm}\right)^3 = 6.4896E4\, MeV/mm^3$$

This conversion is the approximation of x-ray energy absorbed per cubic millimeter of brain tissue. Then, let us approximate the number of nanophosphors in a cubic millimeter of brain tissue. Chen et al. performed *in vivo* imaging studies and the effective nanophosphors concentration used was 133 µg/cm$^3$ [18]. According to the manufacturer, there are about 3.25 x $10^{13}$ nanophosphors per gram or 3.25 x $10^7$ per microgram. Under the assumption that the nanophosphors are distributed at a concentration of 133 µg/cm$^3$ of tissue, a mass ratio can be calculated to be about 1 gram of nanophosphors to 8000 grams of tissue. This ratio will be used to approximate the amount of the absorbed energy that will be converted to visible light photons. The conversion efficiency is dependent on the nanophosphors chosen for a particular application. The $Gd_2O_2S{:}Tb^{3+}$ nanophosphors have an emission wavelength of 545 nm, being close to wavelengths at which halorhodopsin and archeorhodopin are sensitive. Furthermore, Chen et al. have reported that the conversion efficiency of these particles to be about 60,000 visible photons per MeV of absorbed x-ray energy[18]. Using these assumptions the following conversions can be done.

$$6.4896E4\frac{MeV}{mm^3}\left(\frac{1\, g\, NP}{8000\, g\, Tissue}\right)\left(\frac{60000\, photons}{MeV}\right) = 4.8672E5\, photons/mm^3$$

$$4.8672E5\frac{photons}{mm^3}\left(\frac{1000\, mm^3}{133\, \mu g}\right)\left(\frac{\mu g}{3.25E7\, nanophosphors}\right) = \frac{0.11\, photons}{nanophosphor}$$

Less than one photon per nanophosphor is not enough photons to activate ChR2. However, two of the assumptions can be altered to greatly enhance the number of photons per nanophosphor. The first is the phosphor diameter. The nanophosphor mass used in the equation was for 50 nm diameter nanophosphors. Simply by increasing the nanophosphor diameter by a factor of 3 (150 nm), the nanophosphor mass will be increased by 27 times ($3^3$), assuming the material density is constant. This increase boosts emission to 3 photons per nanophosphors. Further improvement can be achieved by increasing the conversion efficiency of the nanoparticles. The current conversion factor (60,000 photons/MeV) is only 15% efficient as there is enough energy to generate ~400,000 visible photons (496 nm) in one MeV. Therefore every 5% increase in efficiency is equal to an increase of 20,000 photons. With both of these adjustments considered, phosphors with a diameter of 150 nm and a quantum efficiency of 50% will emit more than **10 photons per nanophosphor under the acceptable x-ray dose**.

The second part of the feasibility analysis is the number of light photons needed to open the light-activated ion channels. To approximate this number, an understanding of the gating mechanism in the proteins is necessary. The light-sensitive moiety of rhodopsins such as ChR-2 and halorhodopsin, is retinal. Retinal is a covalently bound derivative of Vitamin A that isomerizes under light excitation. According to Hegemann and Möglich, the sensitivity of channelrhodopsin is defined in part by the quantum efficiency of retinal. This is described as the likelihood of the chromophore to isomerize after absorption of a single photon of light. This efficiency falls between 30-70% in rhodopsins[60]. With this in mind, an ion channel will need between **1.5 and 3 photons of absorbed light to isomerize the retinal molecule and trigger an opening in the channel**. As calculated above, by increasing the radius of the particles alone, sufficient numbers of photons can be generated by the delivered x-ray dose to open the ion channels.

### 2.4. U-optogenetics via Sonoluminescence

Sonoluminescence was first discovered in 1934 as a consequence of air bubbles in a photo-developing solution that emitted short bursts of light when subjected to ultrasonic waves[61]. The principle of this effect is that within these air bubbles, there are collisions between free electrons and ions, and when the air bubbles collapse, all these collisions result in thermal bremsstrahlung radiation, which is released as a short burst of light[62]. Studies have shown that the sonoluminescence effect can be enhanced by the introduction of a chemiluminescent agent, such as fluoresceinyl *Cypridina* luminescent analog (FCLA), which reacts with oxygen free radicals in air bubbles[2]. Under ultrasonic waves with a pressure of 200 kPa, FCLA molecules dissolved in water were reported to emit strong chemiluminscence at a peak wavelength of 532 nm with an intensity of 12580 photons cm$^{-2}$ s$^{-1}$ in a mouse model[2]. This characteristic presents an ideal emission wavelength for use in optogenetics, namely u-optogenetics. FCLA would need to be targeted to ion channels in a similar method to what was previously mentioned for x-ray excitable nanophosphors. When subjected to ultrasound waves, the collapsing air bubbles would emit bursts of light to trigger the subsequent stimulation of ion channels that absorb photons at 532 nm, including halorhopsin and archeorhodopsin.

Figure 1A illustrates the use of sonoluminescence to stimulate the ion channels. Under these conditions, sonoluminescence provides an alternative excitation pathway in optogenetics. The advantage of ultrasound over x-ray methods is that no radiation dose would be introduced to the patient. However, there is greater attenuation of ultrasonic waves in tissue and bone as compared to x-rays, so penetration depth would be limited. By using low frequency ultrasound waves, penetration depth can be maximized. For ultrasound waves with a frequency of 1 MHz, the penetration depth in bone is approximately 0.3 cm; at a wave frequency of 100 kHz, the penetration depth would increase to approximately 3 cm[63]. Further, a new study has been reported that may enable even greater penetration depths for ultrasound through the skull by use of acoustic complementary metamaterials that can cancel out aberrating layers in bone[64].

The feasibility of u-optogenetics hinders mainly on the ability of FCLA, or another chemiluminscent agent, to target ion channels directly. Ultrasonic stimulation provides a non-invasive way to stimulate light emission with greater depth than traditional optogenetics. U-optogenetics also has an advantage over x-optogenetics by not delivering radiation, but it does not equal the penetration distance of x-ray techniques.

## 3. Discussions and Conclusions

Putting the pieces together, x-optogenetics seems to be a promising approach beyond the currently accepted optogenetic techniques. Figure 1B illustrates the combination of the key elements needed for x-optogenetics as described in the Methodology and Results section. By replacing the light sources in the form of lasers or LED with x-ray excitable nanophosphors, the issues of invasiveness and depth-limitedness of optogenetic stimulation can be addressed. Through functionalization of the nanoparticles, a desirable targeting capability can be achieved that will allow for accumulation of the nanophosphors near the light-sensitive ion channels. When choosing the nanophosphors, those with high energy conversion efficiency will be preferred as they will work with lower x-ray dose, given the minimum power emission for cell stimulation. Furthermore, size distribution of nanoparticles will also affect the dose needed to achieve sufficient visible light emission. Polycapillary lenses or zone plates can be used to focus x-rays onto altered cells. The x-ray flux will directly affect the density of the emitted light. Through the use of a carbon nano-tube x-ray source rather than a conventional source, a high level of temporal control can be implemented over x-ray excitation, inducing luminescence pulses from the nanophosphors at suitable frequencies and duty cycles.

X-optogenetics is a feasible idea since it uses a safe x-ray dose to excite nanophosphors allowing photon emissions that will be able to reach and activate ion channels in targeted, light-sensitive ion channels. As previously stated, there will be enough photons to activate the ion channels if the radius and/or conversion efficiency of the nanophosphor is increased. It should be noted that increasing the radius of the nanophosphor will lead to additional considerations. For example, the nanophosphor must be small enough to pass through the blood brain barrier (BBB); therefore, increasing the nanophosphor radius could decrease the nanophosphor penetration into the BBB. Optimization of size distribution, particle emission, and BBB penetration will be a key consideration moving forward with x-optogenetics. If it is determined that the nanophosphor's radius is increased by more than what would pass through the BBB, x-optogenetics could also be applied to other regions of the body[65].

Additionally, the importance of targeting x-ray excitable nanophosphors to the ion channels should be noted. The closer the nanophosphors are to the ion channels, the more photons there will be able to activate them. Therefore, the nanophosphors should be targeted to the ion channels as closely as possible. It is likely that only a small number of channels will be directly targeted by the nanophosphors relative to the number expressed in a given cell. This may have a substantial impact on the ability for x-optogenetics to have macroscopic and behavioral effects.

In most optogenetic studies, the light stimuli are delivered in sub-second pulse trains over relatively longer periods. We have discussed the importance of using the CNT for having millisecond control over the x-ray delivery, however, in the feasibility analysis the whole dose is assumed to be used in a single pulse. Clearly, administering a x-ray pulse train of 10 mSv each greatly increases the total effective dose, putting the subject at risk for radiation poisoning. However, a recent study has looked into the inhibitory effects of ChR2-based mutants after a single light pulse[66]. These variants can have effects that outlast the light stimulus. Therefore, x-optogenetics remains a feasible option for these variants since a single x-ray dose resulting in a single light stimulus will cause lasting inhibition in the target neurons.

U-optogenetics via sonoluminescence provides a second alternative method to stimulating ion channels without the need for implanted light sources. This technique differs from x-optogenetics in that it relies on ultrasound waves as the medium for inducing light emission, instead of x-rays, and therefore does not introduce a radiation dose. The penetration depth of u-optogenetic techniques would not be as high as in x-optogenetic methods, but 3 cm penetration depth through the skull using 100 kHz ultrasonic waves would still be a substantial advantage over traditional optogenetics. Sonoluminescence would be enhanced by a chemiluminescent agent such as FCLA, which would emit bursts of light from air bubbles collapsing under ultrasonic pulses. Targeting of FCLA to ion channels provides a means for direct stimulation, but this mechanism remains an area of further investigation. Furthermore, u-optogenetics will not have the pulse-train limitations as radiation dose is not an issue for this technique.

Without the use of a light probe, x-optogenetics and u-optogenetics turn optogenetics into a less invasive and more applicable research tool. The decreased invasiveness puts optogenetics one step closer to being applied to subjects other than rodents. Additionally, it makes optogenetics a less time-consuming and more ethical process since researchers no longer need to surgically drill into the skull of their subjects. Moreover, the ability for x-optogenetics and u-optogenetics to be performed deeply into the tissue would allow researchers to study parts of the brain that current practice of optogenetics does not allow. This would create a grand opportunity to learn and explore parts of the brain that have yet to be explored.

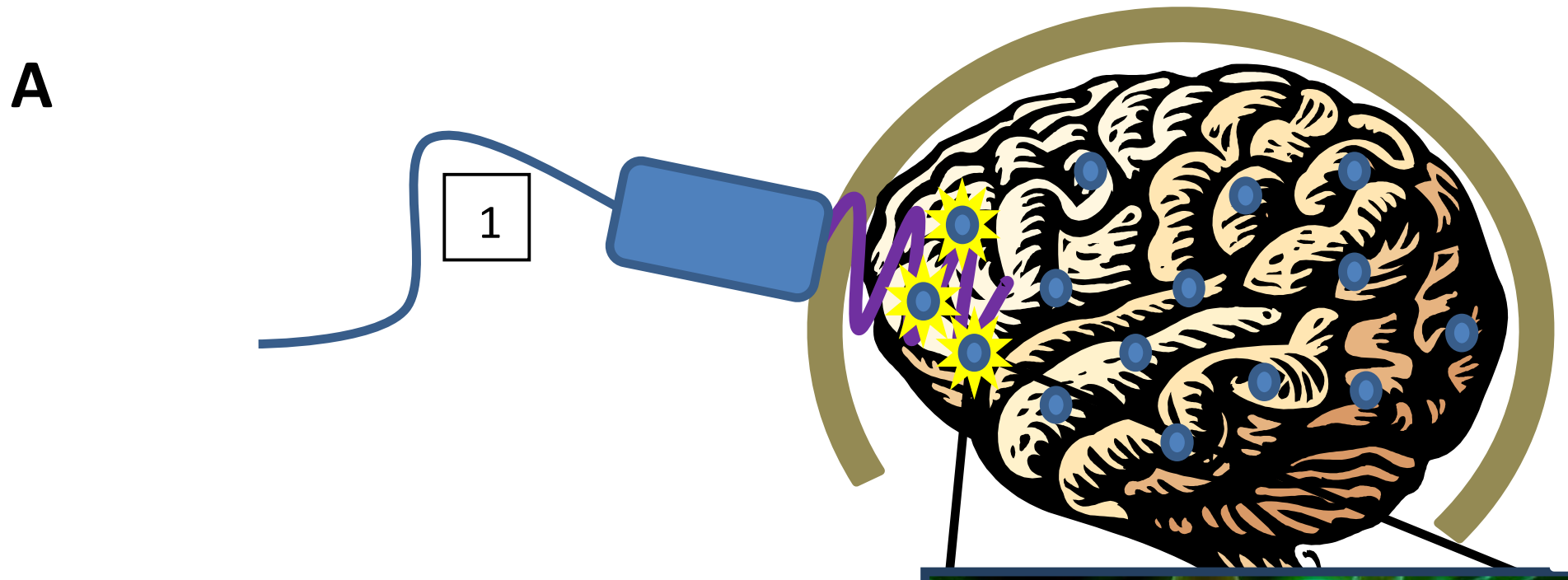

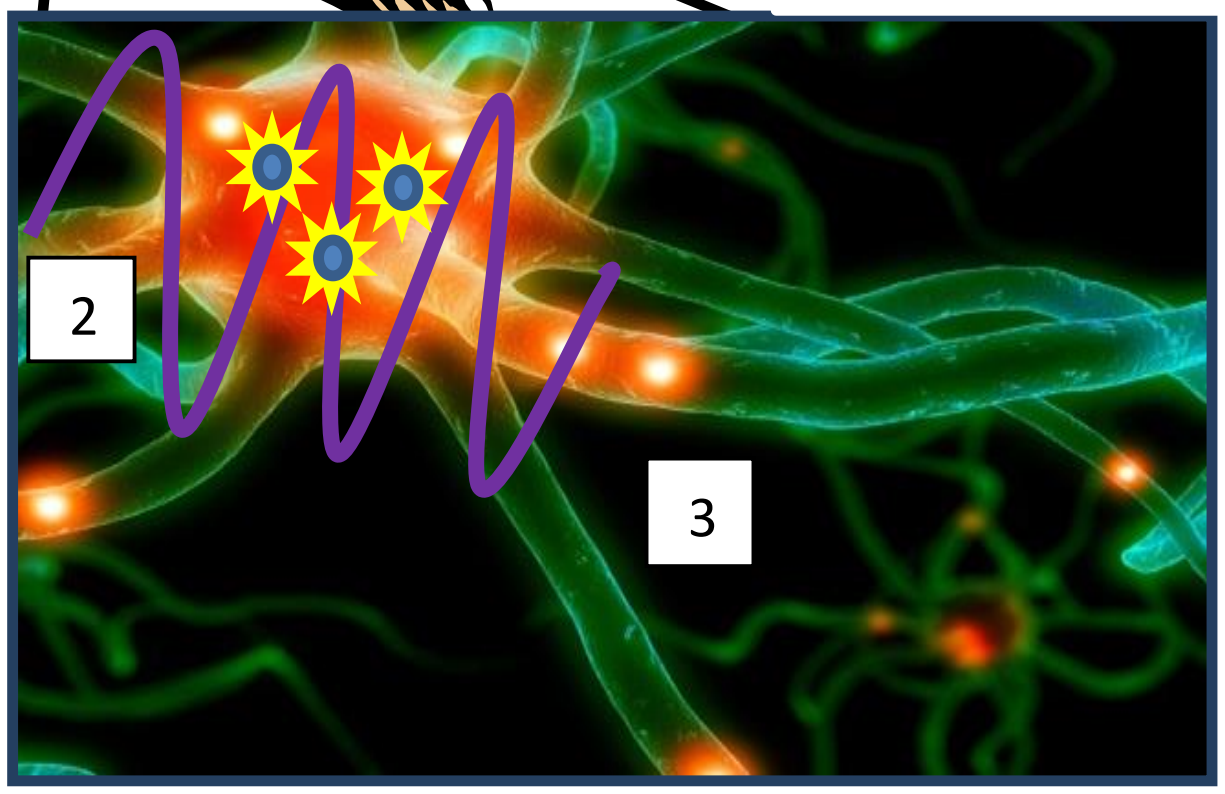

1- Ultrasound transducer delivers kHz to MHz vibrations through the skull which penetrate up to 3 cm.
2- Pressure fluctuations induced by the ultrasonic vibrations will cause the targeted FCLA to enhance sonoluminescence phenomena.
3- Light-sensitive ion channels receiving a single photon of light as a stimuli will open between 30 and 70% of the time and allow for ion flow, changing the membrane potential and possibly stimulating an action potential.

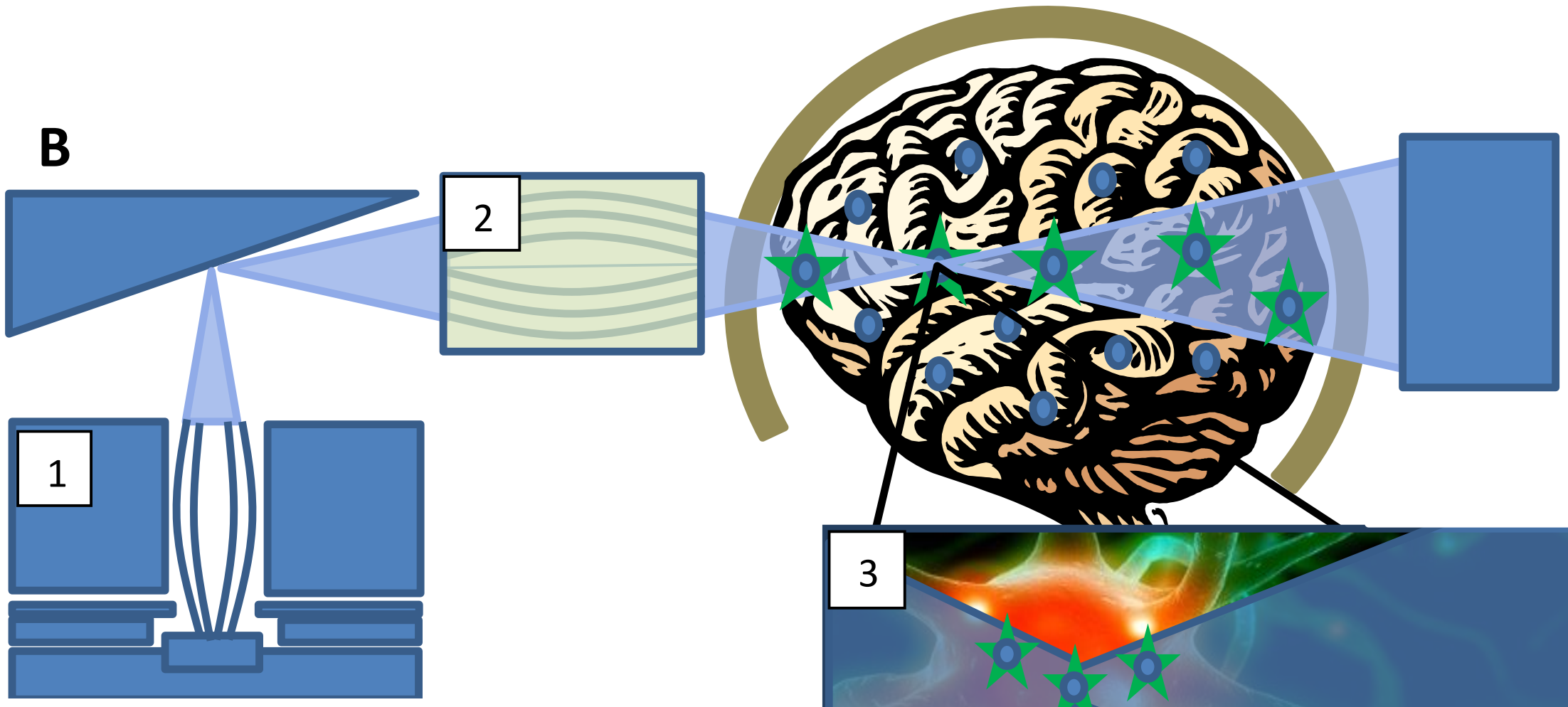

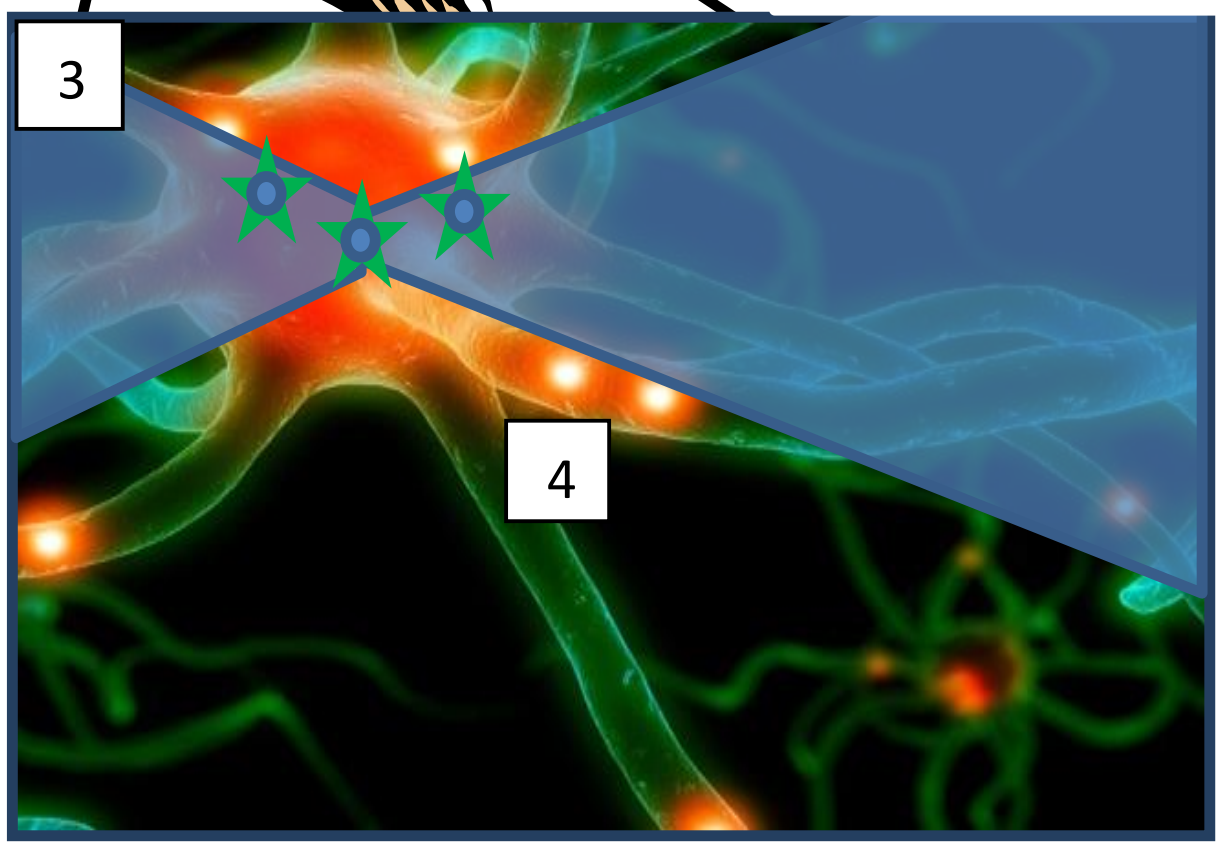

1- Carbon nanotube cathode emits electrons which are focused onto the anode through a gate and focusing electrode.
2- X-rays emitted from the anode are focused into the tissue through a polycapillary lens. The max dose for x-optogenetics is assumed to be 10 mSv. Any unabsorbed radiation will be collected by a detector or x-ray stop.
3- Nanophosphors are tagged to the light sensitive ion channels. Those NPs which are irradiated by the focused x-ray can emit up to 10 photons of light.
4- Light-sensitive ion channels receiving a single photon of light as a stimuli will open between 30 and 70% of the time and allow for ion flow, changing the membrane potential and possibly stimulating an action potential.

Figure 1. A.) Schematic of U-optogenetics that heuristically demonstrates use of ultrasound to induce sonoluminescence and modulate membrane potential. B.) Schematic of X-optogenetics showing the use of a CNT source as well as a polycapillary lens for x-ray focusing into a double-cone geometry. CNT source schematic was adapted from Zhang et al[67]. Brain and neuronal cell images were sourced from Microsoft clipart[68,69].